\documentclass[twocolumn,english]{revtex4-1}
\usepackage[T1]{fontenc}
\usepackage[latin9]{luainputenc}
\usepackage{textcomp}
\usepackage{amsmath}
\usepackage{amssymb}
\usepackage{graphicx}
\usepackage{setspace}

\makeatletter
\usepackage{babel}

\makeatother

\usepackage{babel}
\begin{document}

\title{Hall coefficient and magnetoresistance in boson+fermion dimer models
for the Pseudogap phase of high Tc superconductors}

\author{Garry Goldstein}

\affiliation{TCM Group, Cavendish Laboratory, University of Cambridge, J. J. Thomson
Avenue, Cambridge CB3 0HE, United Kingdom}
\begin{abstract}
We show that the Hall coefficient of the boson+fermion dimer model
for the pseudogap phase of high temperature superconductivity introduced
in \cite{key-30,key-32,key-34,key-2} changes sign from negative at
low temperatures to positive at high temperatures at a characteristic
temperature scale of $\kappa k_{B}T_{0}\sim\hbar\omega_{C}$ (the
cyclotron frequency of the fermionic dimers, here $\kappa\sim0.7\,O(1)$
fits the experimental data well \cite{key-11-1,Leboeuf2011,Ramshaw2015}).
We show that this is an effect of the changing of the sign of the
coupling between the fermionic dimer and the magnetic field from negative
coupling $\sim-e$ at low temperatures to positive coupling $\sim+e$
at high temperature, with the Hall coefficient being proportional
to $R_{H}\sim e_{B}e_{E}e_{J}$ (the product of the magnetic charge,
electric charge and current charge all of which we carefully define).
We relate the Hall conductivity to the coefficient in Kohler's like
rule for magnetoconductivity and calculate some corrections which
are relevant near the intermediate temperature range $\sim50K$ (typical
values for $k_{B}T_{0}\sim\hbar\omega_{C}$). Furthermore we make
a sharp prediction that the magnetoresistance effect vanishes to order
$B^{2}$ at the temperature and magnetic field where the Hall coefficient
vanishes. 
\end{abstract}
\maketitle

\section{\label{sec:Introduction}Introduction}

The pseudogap phase is one of the most enigmatic phases of high temperature
cuprate superconductivity. Recently there has been considerable experimental
evidence that the pseudogap phase of the cuprates has a description
in terms of a vanilla Fermi liquid with nearly free fermionic quasiparticles.
Indeed transport experiments on the cuprates show that the quasiparticle
lifetime $\tau\left(T,\omega\right)$ follows conventional Fermi liquid
behavior with $\tau^{-1}\left(T,\omega\right)\sim\omega^{2}+c^{2}T^{2}$
\cite{key-15} (where $c$ is some constant). Furthermore the pseudogap
phase, at high temperatures $\sim100K-200K$, obeys Kohler's rule
for in plane magnetoresistance with the longitudinal resistance being
proportional to $\rho_{xx}\sim\tau^{-1}\left(1+bH^{2}\tau^{2}\right)$
\cite{key-16} where $b$ is again some constant. Even more evidence
of the existence on nearly free fermionic quasiparticles obeying Fermi
Dirac statistics comes from the observation of quantum oscillations
for the underdoped cuprates \cite{key-17}. The frequency of the oscillations
being between 500 and 600 $T$ showing that there is a very small
Fermi surface with a Fermi area $\sim p/8$ (where $p$ is the doping)
indicating there are $2\times4$ pockets of area $p/8$ each (where
the factor of $2$ comes from spin degeneracy \cite{key-12}), furthermore
the amplitude of these magnetic oscillations follows very well the
Lifshitz-Kosevich formula for the amplitude of quantum oscillations
of free fermions \cite{key-17,key-12}. A description of the pseudogap
phase in terms of nearly free fermions has been achieved in ref. \cite{key-2},
which we follow in this work. A good way to compare the properties
of the Fermi liquid introduced in ref. \cite{key-2} and the Fermi
liquid for the pseodugap is the study of the Hall coefficient, which
is a powerful probe of the Fermi surface of most Fermi liquids. The
sign of the Hall coefficient allows one to determine if the charge
carries are particles or holes \cite{key-10}. One of the most enigmatic
aspects of the underdoped cuprates, the pseudogap phase, is that the
Hall coefficient switches signs from negative to positive as a function
of temperature and magnetic field \cite{key-11-1,Leboeuf2011,Ramshaw2015}.
The transition from negative Hall coefficient at low temperatures
to positive Hall coefficient at high temperatures moves to progressively
higher and higher temperature with increasing magnetic field \cite{key-11-1,Leboeuf2011,Ramshaw2015},
furthermore as we shall see below the transition temperature where
the Hall coefficient goes to zero is proportional to the cyclotron
frequency of quantum oscillations. Any faithful model of the pseudogap
of the high temperature superconductors must reproduce this qualitative
feature. Building on previous work \cite{key-2} this is the main
thrust of this research. We also study the magnetoresistance and show
that it obeys Kohler's rule like effect with a temperature and field
dependent constant $b\left(T,B\right)$, (the field dependence is
a slight deviation from the conventional Kohler's rule behavior of
the magnetoconductivity, though the field dependence may be neglected
at both low and high temperature and is only relevant at an intermediate
temperature range $\sim50K$ or the cyclotron frequency relevant to
that particular doping), we further relate the coefficient $\tilde{b}\left(T,B\right)$
in the magnetoconductivity $\sigma_{xx}^{-1}\sim\tau^{-1}\left(1+\tilde{b}\left(T,B\right)H^{2}\tau^{2}\right)$
to the Hall conductivity $\sigma_{xy}\left(B,T\right)$ and the effective
mass of quantum oscillations $m^{*}$ leading to a relation that involves
only directly measurable quantities \cite{key-17,key-12}.

The Rokhsar-Kivelson quantum dimer model (QDM) was introduced to describe
a magnetically disordered phase (the resonating valence bond (RVB)
phase) of the underdoped cuprate materials \cite{key-9}. Recently
QDMs have been once again revisited as models of high-temperature
superconductivity \cite{key-30,key-32,key-34}. This was motivated
by the need to reconcile transport experiments \cite{key-21} and
photoemission data \cite{key-25} in the underdoped region of cuprate
superconductors. Photoemission data shows Fermi arcs enclosing an
area of $1+p$ (with $p$ being the doping), while transport measurements
indicate plain Fermi-liquid properties consistent with an area $p$.
The authors of Refs. \cite{key-30,key-32,key-34} introduced a model
for the pseudogap region of the cuprate superconductors which consists
of two types of dimers: one spinless bosonic dimer (representing a
valence bond between two neighboring spins) and one spin 1/2 fermionic
dimer representing a hole delocalized between two sites. Using a slave-boson/slave-fermion
approach \cite{key-2}, we were previously able to confirm the numerical
results of refs. \cite{key-30,key-32,key-34} analytically supporting
the existence of a fractionalized Fermi liquid enclosing an area $p$
and to extend this model to show that, in fact, it describes a larger
portion of the phase diagram and captures well the emergence of d-wave
superconductivity \cite{key-2}. Indeed using a meanfield approach
we presented a model of the pseudogap where the fermionic dimers are
nearly free quasiparticles obeying Fermi Dirac statistics which can
condense to form superconductivity \cite{key-2}.

In this work, within the meanfield introduced in reference \cite{key-2}
we will show how the Hall coefficient switches from negative to positive
values as a function of temperature. We will show that the effective
coupling to a static external magnetic field changes sign from $-e$
to $+e$ in a crossover with a transition temperature $T_{0}\left(B\right)$
(where $e_{B}\left(T_{0}\left(B\right),B\right)=0$) given by $\kappa k_{B}T_{0}\left(B\right)\simeq\hbar\omega_{C}=\frac{eB}{m_{f}}$
(the cyclotron frequency of the fermionic dimers \cite{key-17} with
$\kappa\simeq0.7,\,O(1)$ fitting the experimental data) leading to
a change of sign of the Hall coefficient. Indeed we show that $R_{H}\sim e_{B}e_{E}e_{J}$
(the product of the magnetic charge, electric charge and current charge
which we carefully define with $e_{J}=e_{E}=+e$ independent of temperature).
Over all this matches well with experimental data on the underdoped
region of the cuprates \cite{key-11-1,Leboeuf2011,Ramshaw2015} (the
cyclotron frequency is highly doping dependent and its dependence
on doping is well reproduced in the doping dependence of $\kappa T_{0}\left(B\right)$,
see also Appendix \ref{sec:Comparing-with-experiments}). We also
compute explicit formulas for the magnetoresistance and the Hall coefficient
(see Eq. (\ref{eq:Conductivity})). We make a sharp prediction that
the magnetoresistance effect vanishes to order $B^{2}$ when the Hall
resistivity goes through zero, or in other words the coefficient in
Kohler's rule of magnetoresistance vanishes at the temperature and
magnetic field where the Hall coefficient is zero, we further relate
the coefficient $\tilde{b}\left(T,B\right)$ for magnetoconductivity
to the Hall conductivity $\sigma_{xy}$ and the effective mass of
quantum oscillations $m^{*}$ all directly efficiently experimentally
measurable\cite{key-17,key-12}.

In Section \ref{Main-Hamiltonian} we review the form of the main
Hamiltonian used in the text. In section \ref{sec:Qualitative-arguments}
we show how to project the $t-J$ model Hamiltonian with the external
magnetic and electric fields onto the dimer subspace. Section \ref{sec:Hall-Coefficient}
is our main result which shows the Hall coefficient and magnetoresistance
as a function of temperature and field. In the appendices we review
the semiclassical equations of motion needed to derive our key results.

\section{\label{Main-Hamiltonian}Main Hamiltonian}

We will consider a system of dimers as described in reference \cite{key-2}.
The total Hamiltonian for our system is given in \cite{key-2}, we
will also use the notation introduced in ref \cite{key-2}. As pointed
out in reference \cite{key-2} we can make substantial progress in
understanding the fermionic component of the theory without detailed
analysis of the bosonic component. Indeed, any translationally invariant
(liquid-like) ansatz for the bosonic dimers introduced in refs. \cite{key-30,key-32,key-34,key-2}
that does not break the symmetry between the $x$ and $y$ axis, yields
similar fermionic effective theories. The effective fermionic mean-field
Hamiltonian reads \cite{key-2}: 
\begin{align}
H_{F\bar{B}} & =\sum_{\sigma}\sum_{i}\left\{ -t_{1}\;c_{i+\hat{y},\hat{x},\sigma}^{\dagger}c_{i,\hat{x},\sigma}^{\;}\langle b_{i,\hat{x}}^{\dagger}b_{i+\hat{y},\hat{x}}^{\;}\rangle+1\;\text{term}\right.\nonumber \\
 & -t_{1}\;c_{i+\hat{x},\hat{y},\sigma}^{\dagger}c_{i,\hat{y},\sigma}^{\;}\langle b_{i,\hat{y}}^{\dagger}b_{i+\hat{x},\hat{y}}^{\;}\rangle+1\;\text{term}\nonumber \\
 & -t_{2}\;c_{i,\hat{y},\sigma}^{\dagger}c_{i,\hat{x},\sigma}^{\;}\langle b_{i+\hat{x},\hat{y}}^{\dagger}b_{i+\hat{y},\hat{x}}^{\;}\rangle+7\;\text{terms}\nonumber \\
 & -t_{3}\;c_{i,\hat{y},\sigma}^{\dagger}c_{i+\hat{x}+\hat{y},\hat{x},\sigma}^{\;}\langle b_{i+\hat{x}+\hat{y},\hat{x}}^{\dagger}b_{i,\hat{y}}^{\;}\rangle+7\;\text{terms}\nonumber \\
 & \left.-t_{3}\;c_{i,\hat{y},\sigma}^{\dagger}c_{i+2\hat{y},\hat{x},\sigma}^{\;}\langle b_{i+2\hat{y},\hat{x}}^{\dagger}b_{i,\hat{y}}^{\;}\rangle+7\;\text{terms}\right\} \nonumber \\
 & \left(-2\lambda-\mu\right)\sum_{i}\sum_{\sigma}c_{i,\hat{x},\sigma}^{\dagger}c_{i,\hat{x},\sigma}^{\;}\;\nonumber \\
 & \left(-2\lambda-\mu\right)\sum_{i}\sum_{\sigma}c_{i,\hat{y},\sigma}^{\dagger}c_{i,\hat{y},\sigma}^{\;}\;\label{eq:Main_Hamiltonian}
\end{align}
which is effectively a tight-biding model with renormalized hoppings
$T_{1}=t_{1}\;\langle b_{i,\hat{x}}^{\dagger}b_{i+\hat{y},\hat{x}}^{\;}\rangle$,
$T_{2}=t_{2}\;\langle b_{i+\hat{x},\hat{y}}^{\dagger}b_{i+\hat{y},\hat{x}}^{\;}\rangle$
and $T_{3}=t_{3}\;\langle b_{i+\hat{x}+\hat{y},\hat{x}}^{\dagger}b_{i,\hat{y}}^{\;}\rangle$.
Here $c_{i,\eta,\sigma}^{\dagger}$ refers to a fermionic creation
operator (representing a fermionic dimer on the link connecting vertices
$i$ and $i+\eta$ with spin $\sigma$), while $b_{i,\eta}$ refers
to spinless bosonic dimers \cite{key-2}. $\lambda$ is a Lagrange
multiplier used to enforce the constraint that there is exactly one
dimer per site and $\mu$ is the chemical potential of the electrons
in the pseudogap phase. The coefficients $t_{1/2/3}$ were introduced
in refs. \cite{key-30,key-32,key-34}. Furthermore we assume that
there is no time reversal symmetry breaking so that all expectation
values for the bosons are real at zero external field. We note that
changes in the phases of $\langle b_{i,\hat{x}}^{\dagger}b_{i+\hat{y},\hat{x}}^{\;}\rangle$,
$\langle b_{i+\hat{x},\hat{y}}^{\dagger}b_{i+\hat{y},\hat{x}}^{\;}\rangle$
and $\langle b_{i+\hat{x}+\hat{y},\hat{x}}^{\dagger}b_{i,\hat{y}}^{\;}\rangle$
due to the external magnetic field will play a crucial role in the
dynamics of the fermionic dimers, see the discussion below Eq. (\ref{eq:Eqs_motion_E_field})
below.

\begin{figure}
\begin{centering}
\includegraphics[scale=0.25]{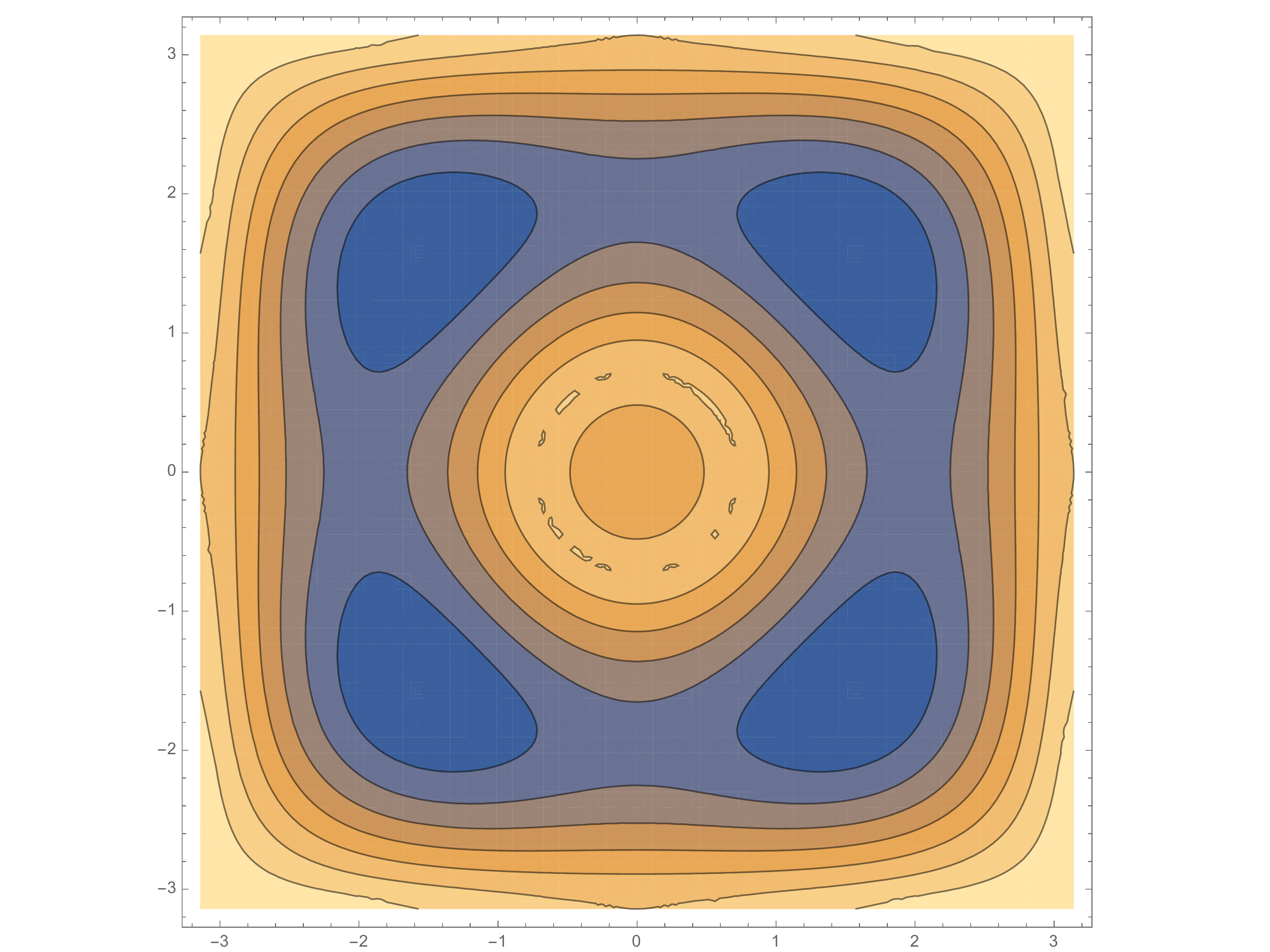} 
\par\end{centering}
\protect\protect \protect\protect\protect

\caption{\label{fig:Typical} A contour energy plot of $E_{-,\vec{k}}$ showing
typical dispersion, at zero electric and magnetic field, with four
pockets near $\left(\pm\frac{\pi}{2},\pm\frac{\pi}{2}\right)$. Details
of the phase diagram of the model are given in ref. \cite{key-2}.}
\end{figure}

The resulting model is defined on a square lattice with a two point
basis. The horizontal ($x$) and vertical ($y$) links make up the
two sublattices where the fermions reside. We define (in momentum
space) the spinor that encodes these two flavors of fermions as $\psi_{\vec{k},\sigma}^{\dagger}=(c_{\vec{k},\hat{y},\sigma}^{\dagger},c_{\vec{k},\hat{x},\sigma}^{\dagger})$
and the Hamiltonian in momentum space is given by \cite{key-2}: 
\begin{equation}
H_{F\bar{B}}=\sum_{\vec{k},\sigma}\psi_{\vec{k},\sigma}^{\dagger}\,
\begin{pmatrix}\xi_{\vec{k}}^{x} & \gamma_{\vec{k}}\\
\gamma_{\vec{k}}^{*} & \xi_{\vec{k}}^{y}
\end{pmatrix}\;\psi_{\vec{k},\sigma}\;,\label{eq:H_k}
\end{equation}
where: 
\begin{align}
\xi_{\vec{k}}^{x} & =-2\lambda-\mu-2\,T_{1}\;\cos k_{x}\;\nonumber \\
\xi_{\vec{k}}^{y} & =-2\lambda-\mu-2\,T_{1}\;\cos k_{y}\;\nonumber \\
\gamma_{\vec{k}} & =4\;\left(T_{2}\;\cos\frac{k_{x}}{2}\cos\frac{k_{y}}{2}\right.\nonumber \\
 & \left.+T_{3}\;\cos\frac{3k_{x}}{2}\cos\frac{k_{y}}{2}+T_{3}\;\cos\frac{k_{x}}{2}\cos\frac{3k_{y}}{2}\right).\label{eq:Definitions}
\end{align}
The eigenvalues are given by $E_{\pm,\vec{k}}=\xi_{\vec{k}}\pm\sqrt{\eta_{\vec{k}}^{2}+|\gamma_{\vec{k}}|^{2}}$,
where $\xi_{\vec{k}}=(\xi_{\vec{k}}^{x}+\xi_{\vec{k}}^{y})/2$ and
$\eta_{\vec{k}}=(\xi_{\vec{k}}^{x}-\xi_{\vec{k}}^{y})/2$. For hole
doping $p$ (the number of fermions in our model) the lower band $E_{-,\vec{k}}$
will be partially occupied \cite{key-2}. The total area enclosed
by the Fermi surface in the lower band is equal to the hole doping
$p/2$. The extra factor of $\frac{1}{2}$ comes from spin degeneracy.
The Hamiltonian Eq.~(\ref{eq:H_k}) has four-fold rotational symmetry,
$k_{x}\rightarrow k_{y}$ and $k_{y}\rightarrow-k_{x}$, and reflection
symmetry about the two axis $k_{x}\rightarrow-k_{x}$ and $k_{y}\rightarrow k_{y}$
as well as $k_{y}\rightarrow-k_{y}$ and $k_{x}\rightarrow k_{x}$.
A typical dispersion showing four Fermi pockets is shown in Fig. (\ref{fig:Typical}).

\section{\label{sec:Qualitative-arguments}Projecting the Hamiltonian }

To study the magnetoresistance and Hall coefficient of the pseudogap
within the boson+fermion dimer model we need to include vector potential
and scalar potential terms in the Hamiltonian, i.e. go beyond the
model introduced in Refs. \cite{key-30,key-32,key-34}. To do so we
start with the the $t-J$ Hamiltonian on the square lattice: 
\begin{equation}
H_{tJ}=-\sum_{\alpha}t_{ij}d_{i,\alpha}^{\dagger}d_{j,\alpha}^{\;}+J\sum_{\left\langle i,j\right\rangle }\left(\vec{S}_{i}\cdot\vec{S}_{j}-\frac{1}{4}n_{i}n_{j}\right)\label{eq:t-J_model}
\end{equation}
subject to the constraint that $n_{i}\leq1$. Here $d_{i,\alpha}^{\dagger}$
and $d_{i,\alpha}^{\ }$ are the electron creation and annihilation
operators ($\alpha=\uparrow,\downarrow$) of the $t-J$ model, $\vec{S}_{i}=\sum_{\alpha,\beta}d_{i,\alpha}^{\dagger}\,\vec{\sigma}_{\alpha,\beta}\,d_{i,\beta}^{\ }$,
and $n_{i}=d_{i,\uparrow}^{\dagger}d_{i,\uparrow}^{\ }+d_{i,\downarrow}^{\dagger}d_{i,\downarrow}^{\ }$.
Under projection, described below, it is not too hard to see that
the term proportional to $J$ does not contribute to the part of the
Hamiltonian that is biquadratic in the fermions and the bosons (see
Eq. (\ref{eq:Main_Hamiltonian}) but only contributes to the fermion
fermion interaction and the boson boson interaction terms in the Hamiltonian
((which leads to superconductivity \cite{key-2} which is suppressed
by the large magnetic field) and produces the RK Hamiltonian \cite{key-9-1}
that acts only on the bosons thereby providing the expectation values
$\langle b_{i,\hat{x}}^{\dagger}b_{i+\hat{y},\hat{x}}^{\;}\rangle$,
$\langle b_{i+\hat{x},\hat{y}}^{\dagger}b_{i+\hat{y},\hat{x}}^{\;}\rangle$
and $\langle b_{i+\hat{x}+\hat{y},\hat{x}}^{\dagger}b_{i,\hat{y}}^{\;}\rangle$)
and therefore will be dropped from now on. We will first include a
vector potential, magnetic field, and scalar potential electric field
into the $t-J$ model using the Pierls substitution: 
\begin{equation}
H_{tJ}=-\sum_{\alpha}t_{ij}\exp\left(i\frac{e}{c}A_{ij}\right)d_{i,\alpha}^{\dagger}d_{j,\alpha}^{\;}-e\sum_{i}\varphi_{i}n_{i}\label{eq:t_J_model_A}
\end{equation}

Where $A_{ij}\equiv\int_{i}^{j}\vec{A}\cdot d\vec{r}$ and $\varphi_{i}$
s the scalar potential at site $i$. We would like to project this
more general Hamiltonian onto the dimer subspace (the projection without
any Pierls substitutions for $E=B=0$ was done in refs. \cite{key-30,key-32,key-34}).
To do so we can identify the dimer Hilbert space with a subspace of
the Hilbert space for the $t-J$ model, where the zero dimers state
(which is outside the physical dimer space) corresponds to the state
with zero electrons, and the rest of the Hilbert space can be introduced
via the operators $b_{i,\eta}^{\dagger}\Leftrightarrow\Upsilon_{i,\eta}\,(d_{i\uparrow}^{\dagger}d_{i+\eta\downarrow}^{\dagger}-d_{i\downarrow}^{\dagger}d_{i+\eta\uparrow}^{\dagger})/\sqrt{2}$
and $c_{i,\eta,\sigma}^{\dagger}\Leftrightarrow\Upsilon_{i,\eta}(d_{i,\sigma}^{\dagger}+d_{i+\eta,\sigma}^{\dagger})/\sqrt{2}$.
The phases $\Upsilon_{i,\eta}$ represent a gauge choice and we shall
follow the one by Rokhsar and Kivelson~\cite{key-9-1} and define
$\Upsilon_{i,\hat{y}}=1$ and $\Upsilon_{i,\hat{x}}=\left(-1\right)^{i_{y}}$,
where $i_{y}$ is the $y$-component of the 2D square lattice site
index $i$. The projection procedure can be described as: 
\begin{equation}
H_{D}=\left(\begin{array}{cc}
\left(\begin{array}{ccc}
\\
 & H_{DD}\\
\\
\end{array}\right) & \begin{array}{cc}
\\
 & H_{DO}\\
\\
\end{array}\\
\begin{array}{ccc}
\\
 & H_{OD}
\end{array} & \begin{array}{cc}
\\
 & H_{OO}
\end{array}
\end{array}\right)\label{eq:Projection_procedure}
\end{equation}
Where we divide the $t-J$ model Hilbert space into the dimer Hilbert
space and its orthogonal complement. We can then write the $t-J$
Hamiltonian in block diagonal form as shown in Eq. (\ref{eq:Projection_procedure})
and keep only the terms $H_{DD}$. We will not do the projection calculation
explicitly but instead we would give qualitative arguments about the
form of the effective Hamiltonian in Section \ref{sec:Hall-Coefficient}.
We will find it convenient to work in the gauge where $E=-\nabla\varphi$
and $B=\nabla\times A$ with $\varphi$ and $A$ time independent.
In this gauge we note that for a time independent electric field the
projection can be carried out directly, and that the electric field
couples to the dimers minimally, e.g. 
\begin{equation}
\varepsilon\rightarrow\varepsilon_{P}+e_{E}\varphi\left(r\right).\label{eq:Minimal_Change}
\end{equation}
Where $E=-\nabla\varphi$ and $\varepsilon_{P}=E_{-}\left(k\right)$
is a periodic Hamiltonian. We note that $e_{E}=+e$ indicating that
under an electric field the dimer acts as a positively charged object
(indeed a bare fermionic dimer has charge $-e$ however when a fermionic
dimer moves a bosonic dimer carrying charge $-2e$ moves in the opposite
direction leading to an electric charge $2e-e=+e$, alternative under
projection the electric field energy is equal to $+e\sum'_{i}\varphi_{i}$
where the sum is taken over the unoccupied electrons (holes) in the
$t-J$ model or equivalently the positions of the fermionic dimers).

\section{\label{sec:Phases-and-Charges}Phases and Charges}

We would like to carefully discuss the phases and charges of the fermionic
and bosonic dimers. There are two relevant gauge groups for the dimers
the internal local $U\left(1\right)$ gauge symmetry \cite{key-2}:
\begin{equation}
b_{i,\eta}\rightarrow e^{i\theta_{i}}\;b_{i,\eta}\;e^{i\theta_{i+\eta}}\;,\quad c_{i,\eta,\sigma}\rightarrow e^{i\theta_{i}}\;c_{i,\eta,\sigma}\;e^{i\theta_{i+\eta}}\;,\label{eq:Gauge_transformation}
\end{equation}
with a phase $\theta_{i}$ associate to each vertex $i$. There is
also the $U\left(1\right)$ due to its coupling to electromagnetism.
We now determine the charges of the bosonic and fermionic dimers under
the electromagnetic gauge field. Under electromagnetism the $t-J$
model electrons transform as 
\begin{equation}
A\rightarrow A-\nabla\alpha,\qquad d_{i}\rightarrow e^{i\frac{e}{c}\alpha}d_{i}\label{eq:Electromagnetism_gauge}
\end{equation}

Now a fermionic dimer is made of a single electron operator while
a bosonic dimer is made of two. This means that under an electromagnetic
gauge transformation the fermionic dimer has charge $-e$ while a
bosonic dimer has charge $-2e$. Indeed or the dimer model we have
the following operator equivalences \cite{key-30}: 
\begin{equation}
b_{i,\eta}^{\dagger}\sim d_{i\uparrow}^{\dagger}d_{i+\eta\downarrow}^{\dagger}-d_{i\downarrow}^{\dagger}d_{i+\eta\uparrow}^{\dagger}\label{eq:dimer_boson_equivalence}
\end{equation}
If $d_{i}\rightarrow e^{i\frac{e}{c}\alpha_{i}}d_{i}$ under a gauge
transformation, then 
\begin{equation}
b_{i,\eta}\rightarrow e^{i\frac{e}{c}\left(\alpha_{i}+\alpha_{i+\eta}\right)},\quad b_{r}\rightarrow e^{2i\frac{e}{c}\alpha\left(r\right)}b_{r}\label{eq:Dimer_boson_phase}
\end{equation}
Therefore $b$ has gauge charge $-2e$ under electromagnetism. Where
we have assumed a long wavelength limit description. Similarly: 
\begin{equation}
c_{i,\eta,\sigma}^{\dagger}\sim d_{i\sigma}^{\dagger}+d_{i+\eta\sigma}^{\dagger}\label{eq:dimer_fermion_equivalence}
\end{equation}
Then for $d_{i}\rightarrow e^{i\frac{e}{c}\alpha_{i}}d_{i}$ under
a gauge transformation 
\begin{equation}
c_{r,\sigma}\rightarrow e^{i\frac{e}{c}\alpha\left(r\right)}c_{r,\sigma}\label{eq:dimer_fermion_phase}
\end{equation}
We now need to calculate the phases $\left\langle b_{i,\eta}^{\dagger}b_{j,\nu}\right\rangle $.
To do so we introduce the electromagnetically gauge invariant greens
functions \cite{Haug1996}:
\begin{equation}
\hat{G}\left(r,r'\right)\equiv\left\langle b^{\dagger}\left(r\right)\exp\left(2i\frac{e}{c}\int_{r}^{r'}A\left(r"\right)dr"\right)b\left(r'\right)\right\rangle \label{eq:Gauge_invarinat}
\end{equation}
These greens functions are gauge invariant \cite{Haug1996}, see also
Appendix \ref{sec:Gauge-Invariance-of}. As such they are rotationally
and translationally invariant (indeed for a constant magnetic field
the system is translationally and rotationally invariant as such a
translation or a rotation is a gauge transformation which does not
change the gauge invariant greens functions), 
\begin{equation}
\hat{G}\left(r,r'\right)=\hat{G}\left(\left|r-r'\right|\right)=\hat{G}\left(r',r\right)=\hat{G}^{*}\left(r,r'\right)\label{eq:Phase_zero}
\end{equation}
As such $\hat{G}\left(r,r'\right)$ has zero phase so 
\begin{equation}
\left\langle b^{\dagger}\left(r\right)b\left(r'\right)\right\rangle \sim\exp\left(-2i\frac{e}{c}\int_{r}^{r'}A\left(r"\right)dr"\right)\label{eq:Greens_functions_phase}
\end{equation}
This derivation however ignores Elitzur's theorem which says that
a gauge symmetry (given in Eq. (\ref{eq:Gauge_transformation}) cannot
be spontaneously broken. Indeed 
\begin{align}
\left\langle b^{\dagger}\left(r\right)b\left(r'\right)\right\rangle  & =0,\label{eq:correlation}\\
\left\langle b^{\dagger}\left(r,\tau\right)b\left(r',\tau\right)b^{\dagger}\left(r',0\right)b\left(r,0\right)\right\rangle  & \sim\exp\left(-T\tau/N\right)
\end{align}
Where $N$ is the number of dimer flavors \cite{Coleman2015,key-2}
($N=1$ for our case). This means that for time scales bigger then
the inverse temperature we have that the dimer expectation value $\left\langle b^{\dagger}\left(r\right)b\left(r'\right)\right\rangle $
has no phase.

\section{\label{sec:Hall-Coefficient}Hall Coefficient and magnetoresistance
(main equations)}

In Appendix \ref{subsec:Main-equations} we obtained that the semiclassical
equations of motion for the fermionic dimers at an arbitrary temperature
in time independent electric and magnetic fields, these are given
by: 
\begin{align}
\dot{k}_{c}\cong & \dot{r}\times e_{B}\left(T,B\right)B+e_{E}\vec{E}\nonumber \\
\dot{r}_{c}\cong & \frac{\partial\varepsilon_{P}}{\partial k_{c}}\label{eq:Eqs_motion_E_field}
\end{align}
Where $\varepsilon_{P}\left(k\right)=E_{-}\left(k\right)$ was introduced
below Eq. (\ref{eq:Definitions}). Furthermore $e_{B}\left(T,B\right)$
depends on temperature and the magnetic field, being negative $-e$
at low temperatures and positive $+e$ at high temperatures with a
crossover temperature given by $\kappa k_{B}T\sim\hbar\omega_{C}=\frac{eB}{m_{f}}\sim50K$
(the cyclotron frequency of the bosonic dimers with mass $m_{B}\sim(1-3)\cdot m_{e}$
at $B\sim50T$ and $\kappa\sim0.7\,O(1)$ fits the experimental data
well, see Appendix \ref{sec:Comparing-with-experiments}). We now
compute $e_{B}\left(T\right)$, the magnetic charge as a function
of temperature and field. First we claim that at zero temperature
$e_{B}=-e$. Indeed when the dimer hops under the effect of the Hamiltonian
in Eq. (\ref{eq:Main_Hamiltonian}) there is charge $+e$ moving with
the dimer. To understand this note that before projection, an electron
of charge $-e$ must move in the opposite direction as the motion
of a fermionic dimer for the dimers to hop. This leads to a contribution
to the phase picked up by a fermionic dimer under hopping of $+e\vec{A}\cdot\triangle\vec{r}/c$
(for a dimer hopping a distance $\triangle r$ coming from the term
$t_{ij}\rightarrow t_{ij}\exp\left(i\frac{e}{c}A_{ij}\right)$). Furthermore
the expectation value $\left\langle b_{i,\eta}^{\dagger}b_{j,\nu}\right\rangle $
contributes at any temperature to the phase which a fermionic dimer
picks up under hopping. At zero temperature it is given by Eq. (\ref{eq:Greens_functions_phase})
or $\left\langle b_{i,\eta}^{\dagger}b_{j,\nu}\right\rangle $ contributes
a phase of $-2e\vec{A}\cdot\triangle\vec{r}/c$ to a fermionic dimer
hopping a distance $\triangle r$ (indeed $T_{1/2/3}\sim$ $\langle b_{i,\hat{x}}^{\dagger}b_{i+\hat{y},\hat{x}}^{\;}\rangle$,
$\langle b_{i+\hat{x},\hat{y}}^{\dagger}b_{i+\hat{y},\hat{x}}^{\;}\rangle$
and $\langle b_{i+\hat{x}+\hat{y},\hat{x}}^{\dagger}b_{i,\hat{y}}^{\;}\rangle$
respectively so pick up the phases of the bosonic expectation values).
This leads to a a total phase for the hopping of a fermionic dimer
of $-2e\vec{A}\cdot\triangle\vec{r}/c+e\vec{A}\cdot\triangle\vec{r}/c=-e\vec{A}\cdot\triangle\vec{r}/c$
so $e_{B}=-e$. We note that this result, that the phase of $\left\langle b_{i,\eta}^{\dagger}b_{j,\nu}\right\rangle $
has a phase of $-2e\vec{A}\cdot\triangle\vec{r}/c$ is only true ignoring
Elitzur's theorem (which can be done in the ground state, zero temperature,
where $\left\langle b_{i,\eta}^{\dagger}b_{j,\nu}\left(\tau\right)b_{j,\nu}^{\dagger}b_{i,\eta}\right\rangle $
is power law correlated). In thermal states $\left\langle b_{i,\eta}^{\dagger}b_{j,\nu}\right\rangle $
does not have a phase for processes longer then the the inverse temperature,
see Eq. (\ref{eq:correlation}). For such process $\left\langle b_{i,\eta}^{\dagger}b_{j,\nu}\right\rangle $
effectively has no phase so the fermionic dimer transform only with
phase $+e\vec{A}\cdot\triangle\vec{r}/c$ leading to positive charge
at large temperature, $e_{B}=+e$. The crossover temperature is given
by the cyclotron frequency of fermionic dimers $\kappa k_{B}T_{0}\sim\hbar\omega_{c}=\frac{eB}{m_{f}}\sim50K$
(the cyclotron frequency at $B\sim50T$ and $\kappa\sim0.7$ fits
the experimental data well, see Appendix \ref{sec:Comparing-with-experiments})
which is the time it takes a for a dimer to go around a fermionic
pocket and as such ``feel'' the magnetic field. By dimensional analysis
$e_{B}\left(B,T\right)=e_{B}\left(\frac{k_{B}T}{\hslash\omega_{C}}\right)$.
Furthermore the fermionic dimers have an effective charge for current
of $e_{J}=+2e-e=+e$, as the motion of a fermionic dimer is anti-correlated
with the motion of a bosonic dimer of charge $-2e$. We note that
this procedure automatically counts the current of the bosonic dimers
so we don't need to add it to the current of the electronic dimers.
We recall that $e_{E}=+e$ indicating that under an electric field
the dimer acts as a positively charged object (since when a dimer
moves a bosonic dimer carrying charge $-2e$ moves in the opposite
direction). Applying the semiclassical equations of motion to the
dimers and obtaining the Boltzmann equation analogously to ref. \cite{key-10},
the Hall response at low temperatures within the relaxation time approximation
for the linearized Boltzmann equation approximation is given by \cite{key-10}:
\begin{align}
\sigma_{xy} & =2\left(e_{E}e_{J}\right)ip\int d^{2}k\cdot m_{x\nu}^{-1}\left(k\right)(k_{\nu}k_{\mu})\left(\frac{\partial f}{\partial\varepsilon}\right)\times\nonumber \\
 & \times\left(\tau^{-1}m_{\nu\beta}\left(k\right)+\frac{e_{B}\left(T,B\right)}{c}\epsilon_{\nu\beta\gamma}B_{\gamma}\right)^{-1}\nonumber \\
 & \cong\sum_{i=1}^{4}\left(\frac{p}{4}\left(e_{E}e_{J}\right)\int d\varepsilon\cdot\varepsilon g^{i}\left(\varepsilon\right)\left(\frac{\partial f}{\partial\varepsilon}\right)\times\right.\nonumber \\
 & \left.\times\left(\tau^{-1}m_{\alpha\beta}^{i}\left(\varepsilon\right)+\frac{e_{B}\left(T,B\right)}{c}\epsilon_{\alpha\beta\gamma}B_{\gamma}\right)^{-1}\right)\times\nonumber \\
 & \left(\int_{-\infty}^{\infty}\varepsilon g^{i}\left(\epsilon\right)\left(\frac{\partial f}{\partial\varepsilon}\right)\right)^{-1}\label{eq:Transport}
\end{align}
Where $g^{i}\left(\varepsilon\right)$ are the density of states for
the four pockets and $m_{\alpha\beta}^{i}\left(k/\varepsilon\right)$
are the local effective masses for the four pockets (and $\varepsilon$
is measured from the bottom of the band). In the last equation we
divided the contribution to the Hall conductivity into four terms
for each of the four fermion pockets see Fig. \ref{fig:Typical}.
For simplicity assuming a uniform effective mass for each pocket with
two principle axis along the $x'$ and $y'$ axis for each of the
four pockets (here by symmetry $x'$ is along the diagonal of the
Brillouin zone going through the pocket and $y'$ is the axis perpendicular
to it) we get that for each pocket \cite{key-10} $\sigma_{\alpha\beta}\left(B\right)=$

\begin{equation}
\frac{p}{4}\frac{e_{E}e_{J}\tau}{\left(1+\left(\frac{e_{B}\left(T,B\right)B}{\sqrt{m_{x'}m_{y'}}}\tau\right)^{2}\right)}\left(\begin{array}{cc}
\frac{1}{m_{x'}} & \frac{e_{B}\left(T,B\right)B}{m_{x'}m_{y'}}\tau\\
-\frac{e_{B}\left(T,B\right)B}{m_{x'}m_{y'}}\tau & \frac{1}{m_{y'}}
\end{array}\right)\label{eq:Matrix_conductivity}
\end{equation}
Where $\alpha,\beta=x',y'$. Now summing over the four pockets and
switching to original co-ordinates we get that \cite{key-10}: 
\begin{align}
\sigma_{xy} & \left(T,B\right)=p\cdot\frac{e_{E}e_{J}\tau}{\left(1+\left(\frac{e_{B}\left(T,B\right)B}{\sqrt{m_{x'}m_{y'}}}\tau\right)^{2}\right)}\frac{e_{B}\left(T,B\right)B}{m_{x'}m_{y'}}\tau\nonumber \\
\sigma_{xx}=\sigma_{yy} & =\frac{p}{2}\frac{e_{E}e_{J}\tau}{\left(1+\left(\frac{e_{B}\left(T,B\right)B}{\sqrt{m_{x'}m_{y'}}}\tau\right)^{2}\right)}\cdot\left(\frac{1}{m_{x'}}+\frac{1}{m_{y'}}\right)\label{eq:Conductivity}
\end{align}
This, Eq. (\ref{eq:Conductivity}) is the main result of this work.
We don't need to count a boson Hall or longitudinal magnetotransport
coefficients since we already counted the motion of the bosons, in
other words when $p=0$ the model introduced in \cite{key-30,key-32,key-34,key-2}
predicts zero conductivity as the bosonic dimers cannot move but merely
exchange positions. In particular the Hall coefficient is negative
for zero temperature since $e_{B}\rightarrow-e$ when $T\rightarrow0$.
In the high temperature limit we have that the dimers couple with
charge $e_{B}=+e$ to the magnetic field, leading to a positive Hall
coefficient with the crossover temperature being given by $k_{B}T\sim h\omega_{C}\sim50K$
see the discussion below Eq. (\ref{eq:Eqs_motion_E_field}).

We now note that the magnetoresistance is given by the matrix: 
\begin{align}
\rho_{\alpha\beta} & =\left(\frac{\left(1+\left(\frac{e_{B}\left(T,B\right)B}{\sqrt{m_{x'}m_{y'}}}\tau\right)^{2}\right)}{pe_{E}e_{J}\tau\left(\left(\frac{e_{B}\left(T,B\right)B}{m_{x'}m_{y'}}\tau\right)^{2}+\frac{1}{4}\left(\frac{1}{m_{x'}}+\frac{1}{m_{y'}}\right)^{2}\right)}\right)\times\nonumber \\
 & \times\left(\begin{array}{cc}
\frac{1}{2}\left(\frac{1}{m_{x'}}+\frac{1}{m_{y'}}\right) & \frac{e_{B}\left(T,B\right)B}{m_{x'}m_{y'}}\tau\\
-\frac{e_{B}\left(T,B\right)B}{m_{x'}m_{y'}}\tau & \frac{1}{2}\left(\frac{1}{m_{x'}}+\frac{1}{m_{y'}}\right)
\end{array}\right)\label{eq:Kohlers}
\end{align}
Where $\alpha,\beta=x,y$. We get that the Kohler's coefficient is
given by: 
\begin{align}
\frac{\rho_{xx}\left(B,T\right)-\rho_{xx}\left(0,T\right)}{\rho_{xx}\left(0,T\right)}\simeq\label{eq:Kohler's_rule}\\
\simeq\left(\left(\frac{e_{B}\left(T,B\right)}{\sqrt{m_{x'}m_{y'}}}\right)^{2}-\left(\frac{2e_{B}\left(T,B\right)}{\left(m_{x'}+m_{y'}\right)}\right)^{2}\right)B^{2}\tau^{2}
\end{align}
The dependence of $e\left(B,T\right)$ on $B$ is a slight deviation
from Kohler's rule but it is only important for intermediate temperatures
$\sim50K$. We note that we can extract the coefficient $\tilde{b}\left(T,B\right)=\left(\frac{e_{B}\left(T,B\right)}{\sqrt{m_{x'}m_{y'}}}\right)^{2}$
of $\sigma_{xx}^{-1}\sim\tau^{-1}\left(1+\tilde{b}\left(T,B\right)H^{2}\tau^{2}\right)$
from the Hall coefficient $\sigma_{xy}\sim\frac{e_{B}\left(T,B\right)}{m_{x'}m_{y'}}$
and the effective mass for quantum oscillations of the cuprates \cite{key-17,key-12}
$m^{*}\cong\sqrt{m_{x'}m_{y'}}$ through the relation: 
\begin{equation}
\sigma_{xy}\left(T,B\right)=p\cdot\frac{e^{2}\tau}{\left(1+b\left(T,B\right)B^{2}\tau^{2}\right)}\frac{\sqrt{\tilde{b}\left(T,B\right)}}{m^{*}}B\tau.\label{eq:Experimental_relation}
\end{equation}
Note that there is a slight deviation from Kohler's law at low temperatures
at $\sim50K$ as $b\left(T,B\right)$ explicitly depends on the magnetic
field through $e_{B}\left(T,B\right)$. Furthermore as a sharp qualitative
test we note that the magnetoresistance effect vanishes when the Hall
conductivity goes to zero as $e_{B}\left(T\right)\rightarrow0$. This
provides a clear test of our theory.

\section{\label{sec:Conclusions}Conclusions}

We have shown that the Hall coefficient of the underdoped cuprates
changes sign as a function of the temperature. We did so by showing
that for static fields the coupling to the magnetic field changes
sign as a function of temperature, while the coupling to the electric
field and the current charge are given by $e_{E}=e_{J}=+e$. The crossover
temperature for the transition between positive and negative $e_{B}$
is given by $\kappa k_{B}T\sim\hbar\omega_{C}\sim50K$ (the cyclotron
frequency of the fermionic dimers at $B\sim50T$ and $\kappa\sim0.7$
fits the experimental data well, see Appendix \ref{sec:Comparing-with-experiments}).
This result matches well with experimental data on the Hall coefficient
of the underdoped cuprates \cite{key-11-1}. This result confirms
further that the model introduced in refs. \cite{key-30,key-32,key-34}
is a good effective model for the pseudogap and that the meanfield
introduced in ref. \cite{key-2} captures most of the qualitative
features of the pseudogap. We also predict that the magnetoresistance
effect vanishes when the Hall conductivity goes to zero, e.g. when
$e_{B}\rightarrow0$. Furthermore we find a relation between the coefficient
in Kohler's like rule for magnetoconductivity and the Hall conductivity
which can be used to further experimentally test the validity of the
theory. We postulate that relation in Eq. (\ref{eq:Experimental_relation})
can be generalized to other materials with quasiparticle descriptions.

\textbf{Acknowledgements:} This work was supported in part by Engineering
and Physical Sciences Research Council (EPSRC) No. EP/M007065/1 and
by the EPSRC Network Plus on ``Emergence and Physics far from Equilibrium\textquotedblright .
Statement of compliance with the EPSRC policy framework on research
data: this publication reports theoretical work that does not require
supporting research data.\textbf{ }The author would like to acknowledge
useful discussions with Claudio Castelnovo, Claudio Chamon and Nigel
Cooper.

\appendix

\section{\label{sec:Comparing-with-experiments}Comparing with experiments}

\begin{figure}
\begin{centering}
\includegraphics[scale=0.25]{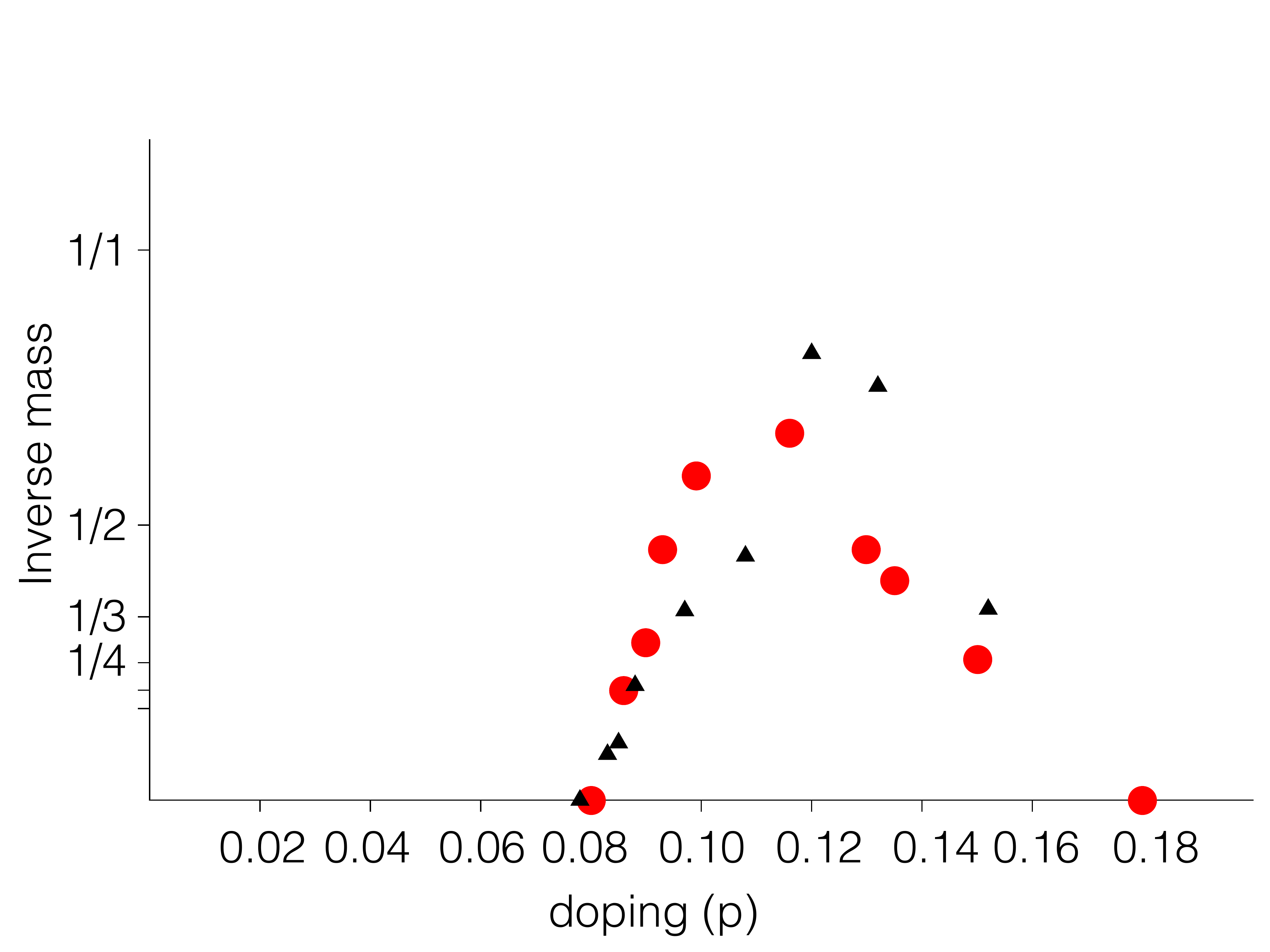} 
\par\end{centering}
\protect\protect \protect\protect\protect

\caption{\label{fig:exoerimental_data} Comparison of the experimental data
for $T_{0}$ and the experimental data for the cyclotron frequency
$\hbar\omega_{C}=\frac{eB}{m_{f}}$. The red triangles represent values
of $\kappa T_{0}/eB$ while the black squares represent $1/m_{f}$,
with $\kappa=0.7$. $1/m_{f}$ is shown in units of $1/m_{e}$. The
black triangles represent $\kappa T_{0}/eB$ \cite{Leboeuf2011} while
the red circles represent $1/m_{f}$ \cite{Ramshaw2015}. }
\end{figure}

The main qualitative output of our work is that $\kappa T_{0}=\hbar\omega_{C}=\frac{eB}{m_{f}}$
for $\kappa=O\left(1\right)$, (here $T_{0}$ is the temperature the
Hall coefficient vanishes). By comparing with the experimental data
we get an excellent fit with $\kappa\cong0.7$ see Fig. (\ref{fig:exoerimental_data}).

\section{\label{sec:Background-on-semiclassical}Background on semiclassical
equations of motion for electrons under general perturbations}

We would like to review the theory of semiclassical electron motion
under general weak slowly time and position dependent perturbations.
this would help us derive Eq. (\ref{eq:Eqs_motion_E_field}) in the
main text, we will closely follow the presentation in refs. \cite{key-11,key-14}.
We will assume that the Hamiltonian can be written as \cite{key-11,key-14}:
\begin{equation}
H\left(r,p;\beta_{1}\left(r,t\right),....\beta_{g}\left(r,t\right)\right)\label{eq:Generic_Hamiltonian}
\end{equation}
Where $\beta_{i}$ are some small perturbations. We will assume that
the fermion is a wave packet centered around the momentum $q_{c}$
and position $r_{c}$. We will assume that the fermion is sufficiently
localized that it is safe to Taylor expand the Hamiltonian \cite{key-11,key-14}:
\begin{align}
H & =H_{c}+\Delta H\nonumber \\
H_{c} & =H\left(r,p;\left\{ \beta_{i}\left(r_{c},t\right)\right\} \right)\nonumber \\
\Delta H & =\frac{1}{2}\sum_{i}\nabla_{r_{c}}\beta_{i}\left(r_{c},t\right)\cdot\left\{ \left(r-r_{c}\right),\frac{\partial H}{\partial\beta_{i}}\right\} \label{eq:Hamiltonian_decomposition}
\end{align}
We see that the Hamiltonian $H_{c}$ has the same periodicity as $H$
as $H_{c}$ is simply shifted by a constant term with respect to $H$.
Therefore it is possible to choose Bloch eigenvalues for the Hamiltonian:
\begin{equation}
H_{c}\left|\psi_{q}\left(r_{c},t\right)\right\rangle =\varepsilon_{c}\left(r_{c},q,t\right)\left|\psi_{q}\left(r_{c},t\right)\right\rangle \label{eq:Eigenvalue_equation}
\end{equation}
Now introducing the Fourier space version of $H_{c}$ we have that
$H_{c}\left(q,r_{c},t\right)=e^{-q\cdot r}H_{c}\left(r_{c},t\right)e^{iq\cdot r}$
and whose eigenstates are the periodic part of the Bloch functions
$\left|u\left(q,r_{c},t\right)\right\rangle =e^{-iq\cdot r}\left|\psi_{q}\left(r_{c},t\right)\right\rangle $.
We then get a Berry potential defined as \cite{key-11,key-14}: 
\begin{equation}
\Lambda_{t,q,r}=\left\langle u\right|\frac{\partial}{\partial t,q,r}\left|u\right\rangle \label{eq:Berry_potential}
\end{equation}
We have the effective Lagrangian \cite{key-11,key-14}: 
\begin{equation}
L=-\varepsilon+q_{c}\cdot\dot{r}_{c}+\dot{q}_{c}\cdot\Lambda_{q}+\dot{r}_{c}\cdot\Lambda_{r}+\Lambda_{t}\label{eq:Lagrangian}
\end{equation}
Here $\varepsilon=\varepsilon_{c}+\Delta\varepsilon$ where 
\begin{equation}
\Delta\varepsilon=\left\langle \Delta H\right\rangle =-Im\left\langle \frac{\partial u}{\partial r_{c}}\right|\cdot\left(\varepsilon_{c}-H_{c}\right)\left|\frac{\partial u}{\partial q}\right\rangle \label{eq:Energy_shift}
\end{equation}
Here the dot product is taken by identifying $r_{c}\in R^{2}$ and
$q_{c}\in R^{2}$. From Euler-Lagrange equations for the Lagrangian
in Eq. (\ref{eq:Lagrangian}) we obtain that \cite{key-11,key-14}:
\begin{align}
\dot{r}_{c} & =\frac{\partial\varepsilon}{\partial q_{c}}-\left(\overleftrightarrow{\Omega}_{q,r}\cdot\dot{r}_{c}+\overleftrightarrow{\Omega}_{q,q}\cdot\dot{q}_{c}\right)-\Omega_{q,t}\nonumber \\
\dot{q}_{c} & =-\frac{\partial\varepsilon}{\partial r_{c}}+\left(\overleftrightarrow{\Omega}_{r,r}\cdot\dot{r}_{c}+\overleftrightarrow{\Omega}_{r,q}\cdot\dot{q}_{c}\right)+\Omega_{q,t}\label{eq:Euler_lagrage_equations}
\end{align}
Where for example 
\begin{equation}
\left(\overleftrightarrow{\Omega}_{q,r}\right)_{\alpha,\beta}=\partial_{q_{\alpha}}\Lambda_{r_{\beta}}-\partial_{q_{\beta}}\Lambda_{r_{\alpha}}\label{eq:Berry_Curvature}
\end{equation}

\section{\label{sec:Equations-of-Motion}Equations of Motion}

\subsection{\label{sec:Computing-the-Berry}Simplifying the Berry curvatures}

We now specialize to the dimer model used in the main text we will
assume that the magnetic and electric fields don't depend on time,
e.g. $\vec{\Omega}_{qt}=\vec{\Omega}_{rt}=0$. We would like to simplify
the Berry curvatures that enter Eq. (\ref{eq:Euler_lagrage_equations})
above for the dimer system. The key formula we will use is that for
a two level system, with a Hamiltonian of the form $n\left(\vec{x}\right)\cdot\sigma+\epsilon\left(x\right)\cdot Id$,
for the lower band the Berry curvature is given by:

\begin{equation}
\Omega_{x_{i}x_{j}}=-\frac{1}{2}\hat{n}\cdot\left(\partial_{x_{i}}\hat{n}\times\partial_{x_{j}}\hat{n}\right)\label{eq:Berry}
\end{equation}
Furthermore we have that 
\begin{equation}
\Delta\varepsilon=\left|\vec{n}\right|\sum_{i=1}^{2}\Omega_{r_{i}k_{i}}.\label{eq:Energy_shift-1}
\end{equation}
Furthermore the Hamiltonian for the dimers, with no magnetic or electric
field, has both time reversal symmetry and inversion symmetry, this
means that \cite{key-14} the berry curvature $\overleftrightarrow{\Omega}_{q,q}$
vanishes identically.

\subsection{\label{subsec:Main-equations}Main equations}

With these simplifications, noting that the Hamiltonian is independent
of time, Eq. (\ref{eq:Euler_lagrage_equations}) simplifies to (we
would like to note again that the changes to the expectation values
of $\left\langle b_{i,\eta}^{\dagger}b_{j,\nu}\right\rangle $ due
to the magnetic field have to be incorporated into this Hamiltonian):

\begin{align}
\dot{r}_{c} & =\frac{\partial\varepsilon_{p}}{\partial q_{c}}-\overleftrightarrow{\Omega}_{q,r}\cdot\dot{r}_{c}\nonumber \\
\dot{q}_{c} & =-\frac{\partial\varepsilon}{\partial r_{c}}+\left(\overleftrightarrow{\Omega}_{r,r}\cdot\dot{r}_{c}+\overleftrightarrow{\Omega}_{r,q}\cdot\dot{q}_{c}\right)\label{eq:Euler_lagrage_equations_simpler}
\end{align}
Furthermore while working in the Landau gauge using Eq. (\ref{eq:Berry})
and noticing that there is no $y$ dependence in the field $\hat{n}$
we obtain that $\overleftrightarrow{\Omega}_{r,r}=0$, since this
is a gauge invariant quantity it vanishes in all gauges, therefore
the equations simplify to: 
\begin{align}
\left(1+\overleftrightarrow{\Omega}_{q,r}\right)\dot{r}_{c} & =\frac{\partial\varepsilon}{\partial q_{c}}\nonumber \\
\left(1+\overleftrightarrow{\Omega}_{q,r}\right)\dot{q}_{c} & =-\frac{\partial\varepsilon}{\partial r_{c}}\label{eq:Euler_lagrage_equations_simpler-1}
\end{align}

Now if we introduce $k_{c}=q_{c}+\frac{e_{B}}{c}A_{c}$ and $\tilde{r}_{c}=r_{c}$
the energy functional simplifies $\varepsilon_{p}\left(r_{c},q_{c}\right)\rightarrow\varepsilon_{p}\left(k_{c}\right)$
with no $\tilde{r}_{c}$ dependence. Now as a function of temperature
or for $e_{B}\left(T,B\right)$ : 
\begin{align}
\frac{\partial\varepsilon}{\partial q_{c}} & =\frac{\partial\varepsilon}{\partial k_{c}}\frac{\partial k_{c}}{\partial q_{c}}=\frac{\partial\varepsilon}{\partial k_{c}}\nonumber \\
\frac{\partial\varepsilon_{p}}{\partial r_{c}} & =\frac{\partial\varepsilon_{P}}{\partial k_{c}}\frac{\partial k_{c}}{\partial r_{c}}=\frac{1}{2}\frac{e_{B}}{c}\frac{\partial\varepsilon_{P}}{\partial k_{c}}\times B\label{eq:Eq_Motion}
\end{align}

Where $B$ can depend on position and we switched to symmetric gauge.
Furthermore 
\begin{equation}
\dot{k}_{c}=\dot{q}_{c}-\frac{1}{2}\frac{e_{B}}{c}\dot{r}_{c}\times B\label{eq:k_c_derivative}
\end{equation}

We get that: 
\begin{align}
\left(1+\overleftrightarrow{\Omega}_{q,r}\right)\dot{r}_{c} & =\frac{\partial\varepsilon}{\partial k_{c}}\nonumber \\
\left(1+\overleftrightarrow{\Omega}_{q,r}\right) & \left(\dot{k}_{c}-\frac{1}{2}\frac{e}{c}\dot{r}_{c}\times B\right)=\nonumber \\
 & =+\frac{1}{2}\frac{e_{B}}{c}\frac{\partial\varepsilon_{P}}{\partial k_{c}}\times B-e_{E}E\nonumber \\
 & =+\frac{1}{2}\frac{e}{c}\left(1+\overleftrightarrow{\Omega}_{q,r}\right)\dot{r}_{c}\times B+e_{E}E\label{eq:Main_Eqs_E_B_field}
\end{align}
The extra $e_{E}E$ term comes from the transformation in Eq. (\ref{eq:Minimal_Change}).
This simplifies to: 
\begin{align}
\left(1+\overleftrightarrow{\Omega}_{k_{c},r}\left(k_{c}\right)\right)\dot{r}_{c} & =\frac{\partial\varepsilon}{\partial k_{c}}\nonumber \\
\dot{k}_{c}-\frac{e_{B}}{c}\dot{r}_{c}\times B-e_{E}E & =0\label{eq:Euler_lagrage_equations_simpler-3}
\end{align}
Where we have ignored a term $\overleftrightarrow{\Omega}_{qr}$ which
is small for small $B$. We have dropped the difference between $q$
and $k_{c}$ in $\overleftrightarrow{\Omega}_{q,r}$ and then dropped
the $r_{c}$ dependence which is zero by gauge invariance. Which further
simplifies to a single equation: 
\begin{equation}
\dot{k}_{c}=\left(1+\overleftrightarrow{\Omega}_{k_{c},r}\left(k_{c}\right)\right)^{-1}\left(\dot{r}_{c}\times e_{B}B+e_{E}E\right)\label{eq:k_c_one_line}
\end{equation}

Furthermore for small $B$ we have that $\overleftrightarrow{\Omega}_{k_{c},r}\left(k_{c}\right)\cong0$
and 
\begin{align}
\dot{k}_{c}\cong & \dot{r}_{c}\times e_{B}B+e_{E}E\nonumber \\
\dot{r}_{c}\cong & \frac{\partial\varepsilon_{P}}{\partial k_{c}}\label{eq:Eq_motion_B_field}
\end{align}

Within the same approximation $\bigtriangleup\varepsilon\cong0$ see
Eq. (\ref{eq:Energy_shift-1}). Where as discussed previously below
Eq. (\ref{eq:Euler_lagrage_equations_simpler-1}) $e_{B}\left(T,B\right)$
depends on temperature and magnetic field. As such we obtain Eq. (\ref{eq:Eqs_motion_E_field})
in the main text. 

\section{\label{sec:Gauge-Invariance-of}Gauge Invariance of the gauge invariant
green's functions}

Under a gauge transformation the gauge invariant greens functions
do not transform, indeed under a gauge transformation in Eq (\ref{eq:Electromagnetism_gauge})
we have that:
\begin{widetext}
\begin{align}
\left\langle b^{\dagger}\left(r\right)\exp\left(2ie\int_{r}^{r'}A\left(r"\right)dr"\right)b\left(r'\right)\right\rangle  & \rightarrow\left\langle b^{\dagger}\left(r\right)e^{-2i\frac{e}{c}\alpha\left(r\right)}\exp\left(2i\frac{e}{c}\int_{r}^{r'}\left\{ A\left(r"\right)-\nabla\alpha\left(r"\right)\right\} dr"\right)b\left(r'\right)e^{2i\frac{e}{c}\alpha\left(r'\right)}\right\rangle \nonumber \\
 & =\left\langle b^{\dagger}\left(r\right)\exp\left(2i\frac{e}{c}\int_{r}^{r'}A\left(r"\right)dr"\right)b\left(r'\right)\right\rangle \label{eq:Gauge_transformation-1}
\end{align}
\end{widetext}


\begin{thebibliography}{10}
\bibitem[1]{key-30} M. Punk, A. Allais and S. Sachdev, PNAS \textbf{112},
9552 (2015).

\bibitem[2]{key-32} D. Chowdhury and S. Sachdev \textit{The enigma
of the pseudogap phase in the cuprate superconductors} in \textit{Quantum
criticality in condensed matter: phenomena, materials and ideas in
theory and experiment} J. Jedrzejewski eds. (Word scientific publishing
co, Singapore 2016).

\bibitem[3]{key-34} A. A. Patel, D. Chowdhury, A. Allais, and S.
Sachdev, Phys. Rev. B \textbf{93}, 165139 (2016)

\bibitem[4]{key-11}G. Sundaram and Q. Niu, Phys. Rev. B \textbf{59},
14915 (1999).

\bibitem[5]{key-14} D. Xiao, M.-C. Chang and Q. Niu, Rev. Mod Phys.
\textbf{82}, 1959 (2010).

\bibitem[6]{key-2} G. Goldstein, C. Chamon and C. Castelnovo, Phys.
Rev. B \textbf{95}, 174511 (2017).

\bibitem[7]{key-15} S. I. Mirzaei, D. Stricker, J. N. Hancock, C.
Berthod, A. Georges, E. van Heumen, M. K. Chan, X. Zhao, Y. Li, M.
Greven, N. Bari\v{ }si\textasciiacute c, and D. van der Marel, Proc.
Nat. Acad. Sci. \textbf{110}, 5774 (2013).

\bibitem[8]{key-16}M. K. Chan, M. J. Veit, C. J. Dorow, Y. Ge, Y.
Li, W. Tabis, Y. Tang, X. Zhao, N. Bari\v{ }si\textasciiacute c, and
M. Greven, Phys. Rev. Lett. \textbf{113}, 177005 (2014).

\bibitem[9]{key-17} S. E, Sebastain, N. Harrison and G. G. Lonzarich,
Phyl. Trans. of Royal Soc. A \textbf{369}, 1687 (2011).

\bibitem[10]{key-9-1} D. S. Rokhsar and S. A. Kivelson, Phys. Rev.
Lett. 61, 2376 (1988).

\begin{spacing}{0.7}
\bibitem[11]{key-16-1} E. Fradkin, D. A. Huse, R. Moessner, V. Ognasian
and S. L. Sondhi, Phys. Rev. B 69, 224415 (2004); A. Vishwanath, L.
Balents and T. Senthil, Phys. Rev. B 69, 224415 (2004).

\bibitem[12]{key-21} Y. Ando, Y. Kurita, S. Komiya, S. Ono, and K.
Segawa, Phys. Rev. Lett. 92, 197001 (2004). J. Orenstein, G. A. Thomas,
A. J. Millis, S. L. Cooper, D. H. Rapkine, T. Timusk, L. F. Schneemeyer,
and J. V. Waszczak, Phys. Rev. B 42, 6342 (1990).

\bibitem[13]{key-25} P. A. Lee, N. Nagaosa and X. G. Wen, Rev. Mod.
Phys. 78 (2006). 
\end{spacing}

\bibitem[14]{key-9} A. Allais, J. Bauer and S. Sachdev, Indian J.
Phys. \textbf{88}, 905 (2014).

\bibitem[15]{key-10} D. Arovas, \textit{Lecture Notes on Condensed
Matter Physics (A Work in Progress)}, (no publisher).

\bibitem[16]{key-11-1} D. LeBoeuf, N. Doiron-Leyraud, J. Levallois,
R. Daou, J.-B. Bonnemaison, N. E. Hussey, L. Balicas, B. J. Ramshaw,
R. Liang, D. A. Bonn, W. N. Hardy, S. Adachi, Cyril Proust and L.
Taillefer, Nature \textbf{450}, 533 (2007).

\bibitem[17]{key-12} G. Goldstein, N. Cooper, C. Chamon, C. Castelnovo,
in preparation.

\bibitem[18]{Haug1996}H. Haug and A.-P. Jauho, Quantum kinetics in
optical and transport in semiconductors, (Springer, Heidelberg, 1996).

\bibitem[19]{Coleman2015} P. Coleman, Introduction to Many-Body Physics
(Cambridge University Press, Cambridge, 2015). 

\bibitem[20]{Ramshaw2015} B. J. Ranshaw, S. E. Sebastian, R. D. McDonald,
J. Bay, B. S. Tan, Z. Zhu, J. B. Betts, R. Liang, D. A. Bonn, W. N.
Hardy and N. Harrison, Science \textbf{348}, 317 (2015).

\bibitem[21]{Leboeuf2011}D. Leboef, N. Doiron-Leyraud, B. Vignolle,
M. Sutherland, B. J. Ramshaw, J. Levallois, R. Daou, F. Laliberte,
O. Cyr-Choinier, J. Chang, Y. J. Jo, L. Balicas, R. Liang, D. A. Bonnm
W. N. Hardy, C. Proust, and L. Taillefer, Phys Rev B \textbf{83},
054506 (2011).
\end{thebibliography}
\end{document}